\documentclass{llncs} 
\usepackage{amsmath}
\usepackage{amssymb} 
\usepackage{pstricks} 
\usepackage{complexity}
\usepackage{tikz}
\usetikzlibrary{shapes,arrows,automata}
\usepackage{array}
\usepackage[margin=3.7cm]{geometry}

\def\Nset{\mathbb{N}} 

\title{On the computational complexity of spiking neural P systems}

\author{Turlough Neary\thanks{The author is funded by Science Foundation Ireland Research Frontiers Programme grant number 07/RFP/CSMFz1.}}
\institute{Boole Centre for Research in Informatics,\\
 University  College Cork, Ireland.\\
\email{tneary@cs.may.ie}}
\begin{document}
\maketitle
\begin{abstract}
It is shown that there is no standard spiking neural P system that simulates Turing machines with less than exponential time and space overheads. The spiking neural P systems considered here have a constant number of neurons that is independent of the input length. Following this we construct a universal spiking neural P system with exhaustive use of rules that simulates Turing machines in linear time and has only 10 neurons. 
\end{abstract}

\section{Introduction}
Since their inception inside of the last decade P systems~\cite{Paun2002} have spawned a variety of hybrid systems. One such hybrid, that of spiking neural P systems~\cite{Ionescu2006}, results from a fusion with spiking neural networks. It has been shown that these systems are computationally universal. Here the time/space computational complexity of spiking neural P systems is examined. We begin by showing that counter machines simulate standard spiking neural P systems with linear time and space overheads. Fischer et al.~\cite{Fischer1968} have previously shown that counter machines require exponential time and space to simulate Turing machines. Thus it immediately follows that there is no spiking neural P system that simulates Turing machines with less than exponential time and space overheads. These results are for spiking neural P systems that have a constant number of neurons independent of the input length.

Extended spiking neural P systems with exhaustive use of rules were proved computationally universal in~\cite{Ionescu2007B}. Zhang et al.~\cite{Zhang2008A} gave a small universal spiking neural P system with exhaustive use of rules (without delay) that has 125 neurons. The technique used to prove universality in~\cite{Ionescu2007B} and~\cite{Zhang2008A} involved simulation of counter machines and thus suffers from an exponential time overhead when simulating Turing machines. In an earlier version~\cite{Neary2008A} of the work we present here, we gave an extended spiking neural P system with exhaustive use of rules that simulates Turing machines in \emph{polynomial time} and has \emph{18 neurons}. Here we improve on this result to give an extended spiking neural P system with exhaustive use of rules that simulates Turing machines in \emph{linear time} and has only \emph{10 neurons}. 

The brief history of small universal spiking neural P systems is given in Table~\ref{tab:Small_SNP}. Note that, to simulate an arbitrary Turing machine that computes in time $t$, all of the small universal spiking neural P systems prior to our results require time that is exponential in $t$. An arbitrary Turing machine that uses space of $s$ is simulated by the universal systems given in~\cite{Ionescu2007B,Neary2008B,Zhang2008A} in space that is doubly exponential in $s$, and by the universal systems given in~\cite{Ionescu2006,Neary2008A,Paun2007,Zhang2008B} in space that is exponential in $s$. 

\begin{table}[h]
\begin{center}
\begin{tabular}{@{}c@{\:}|@{\:}c@{\:}|@{\;}c@{\;}|@{\;}c@{\,}|@{\,}c@{}}
number of & simulation & type  & exhaustive  & author \\
neurons & time/space &of rules& use of rules &\\ \hline
84	& exponential	& standard & no	& P{\u{a}}un and P{\u{a}}un~\cite{Paun2007}\\
67	& exponential	& standard & no & Zhang et al.~\cite{Zhang2008B}\\
49	& exponential	& extended\dag & no & P{\u{a}}un and P{\u{a}}un~\cite{Paun2007}\\
41	& exponential	& extended\dag & no & Zhang et al.~\cite{Zhang2008B}\\ 
12	& double-exponential	& extended\dag & no  & Neary~\cite{Neary2008B}\\
18	& exponential	& extended & no & Neary~\cite{Neary2008B,Neary2008C}*\\
17	& exponential	& standard\dag & no	& ~\cite{Neary2009}\\
14	& double-exponential	& standard\dag & no	& ~\cite{Neary2009}\\ 
5	& exponential	& extended\dag & no	& ~\cite{Neary2009}\\
4	& double-exponential	& extended\dag & no	& ~\cite{Neary2009}\\
3	& double-exponential	& extended\ddag & no	& ~\cite{Neary2009}\\
125	& exponential/	& extended\dag & yes & Zhang et al.~\cite{Zhang2008A}\\
	& double-exponential	&  &  & \\
18	& polynomial/exponential	& extended & yes & Neary~\cite{Neary2008A}\\
\textbf{10}	& \textbf{linear/exponential}	& \textbf{extended} & \textbf{yes} &  \textbf{Section~\ref{sec:small time efficient SNP system}}\\ \hline
\end{tabular}
\end{center}
\caption{Small universal SN P systems. The ``simulation time'' column gives the overheads used by each system we simulating a standard single tape Turing machine. \dag~indicates that there is a restriction of the rules as delay is not used and \ddag~indicates that a more generalised output technique is used. *The 18 neuron system is not explicitly given in~\cite{Neary2008B}; it is however mentioned at the end of the paper and is easily derived from the other system presented in~\cite{Neary2008B}. Also, its operation and its graph were presented in~\cite{Neary2008C}.}\label{tab:Small_SNP}
\end{table}

Chen et al.~\cite{Chen2006A} have shown that with exponential pre-computed resources \textsc{sat} is solvable in constant time with spiking neural P systems. Leporati et al.~\cite{Leporati2007A} gave a semi-uniform family of extended spiking neural P systems that solve the \textsc{Subset Sum} problem in constant time. In later work, Leporati et al.~\cite{Leporati2007B} gave a uniform family of maximally parallel spiking neural P systems with more general rules that solve the \textsc{Subset Sum} problem in polynomial time. All the above solutions to NP-hard problems rely on families of spiking neural P systems. Specifically, the size of the problem instance determines the number of neurons in the spiking neural P system that solves that particular instance. This is similar to solving problems with uniform circuits families where each input size has a specific circuit that solves it. Ionescu and Drago\c{s}~\cite{Ionescu2007A} have shown that spiking neural P systems simulate circuits in linear time.

In the next two sections we give definitions for spiking neural P systems and counter machines and explain the operation of both. Following this, in Section~\ref{sec:Counter machines simulate P systems in linear time}, we prove that counter machines simulate spiking neural P systems in linear time. Thus proving that there exists no universal spiking neural P system that simulates Turing machines in less than exponential time. In Section~\ref{sec:small time efficient SNP system} we present our universal spiking neural P system, with exhaustive use of rules, that simulates Turing machine in linear time and has only 10 neurons. Finally, we end the paper with some discussion and conclusions.

\section{Spiking neural P systems}\label{sec:Spiking neural P-systems}

\begin{definition}[Spiking neural P systems]\label{def:Spiking neural P-systems}
A spiking neural P system is a tuple $\Pi=(O,\sigma_1,\sigma_2,\cdots,\sigma_m,syn,in,out)$, where:
\begin{enumerate}
 \item $O=\{s\}$ is the unary alphabet ($s$ is known as a spike),
 \item $\sigma_1,\sigma_2,\cdots,\sigma_m$ are neurons, of the form $\sigma_i=(n_i,R_i),1\leqslant i\leqslant m$, where:
\begin{enumerate}
 \item $n_i\geqslant 0$ is the initial number of spikes contained in $\sigma_i$,
 \item $R_i$ is a finite set of rules of the following two forms:
\begin{enumerate}
 \item $E/s^b\rightarrow s;d$, where $E$ is a regular expression over $s$, $b\geqslant 1$ and $d\geqslant 1$, 
 \item $s^e\rightarrow\lambda;0$ where $\lambda$ is the empty word, $e\geqslant 1$, and for all $E/s^b\rightarrow s;d$ from $R_i$ $s^e\notin L(E)$ where $L(E)$ is the language defined by $E$, 
\end{enumerate}
\end{enumerate}
 \item $syn\subseteq \{1,2,\cdots,m\}\times\{1,2,\cdots,m\}$ are the set of synapses between neurons, where $i\neq j$ for all  $(i,j)\in syn$, 
 \item $in,out\in\{\sigma_1,\sigma_2,\cdots,\sigma_m\}$ are the input and output neurons respectively.
\end{enumerate}
\end{definition}
In the same manner as in~\cite{Paun2007}, spikes are introduced into the system from the environment by reading in a binary sequence (or word) $w\in\{0,1\}^\ast$ via the input neuron $\sigma_1$. The sequence $w$ is read from left to right one symbol at each timestep. If the read symbol is 1 then a spike enters the input neuron on that timestep.

A firing rule $r=E/s^b\rightarrow s;d$ is applicable in a neuron $\sigma_i$ if there are $j\geqslant b$ spikes in $\sigma_i$ and $s^j\in L(E)$ where $L(E)$ is the set of words defined by the regular expression $E$. If, at time $t$, rule $r$ is executed then $b$ spikes are removed from the neuron, and at time $t+d-1$ the neuron fires. When a neuron $\sigma_i$ fires a spike is sent to each neuron $\sigma_j$ for every synapse $(i,j)$ in $\Pi$. Also, the neuron $\sigma_i$ remains closed and does not receive spikes until time $t+d-1$ and no other rule may execute in $\sigma_i$ until time $t+d$. We note here that in 2b(i) it is standard to have a $d\geqslant0$. However, we have $d\geqslant1$ as it simplifies explanations throughout the paper. This does not effect the operation as the neuron fires at time $t+d-1$ instead of $t+d$. A forgeting rule $r'=s^e\rightarrow\lambda;0$ is applicable in a neuron $\sigma_i$ if there are exactly $e$ spikes in $\sigma_i$. If $r'$ is executed then $e$ spikes are removed from the neuron. At each timestep $t$ a rule must be applied in each neuron if there is one or more applicable rules at time $t$. Thus while the application of rules in each individual neuron is sequential the neurons operate in parallel with each other.

Note from 2b(i) of Definition~\ref{def:Spiking neural P-systems} that there may be two rules of the form $E/s^{b}\rightarrow s;d$, that are applicable in a single neuron at a given time. If this is the case then the next rule to execute is chosen non-deterministically. The output is the time between the first and second spike in the output neuron $\sigma_m$.

An extended spiking neural P system~\cite{Paun2007} has more general rules of the form $E/s^b\rightarrow s^p;d$, where $b\geqslant p\geqslant 0$. Note if $p=0$ then $E/s^b\rightarrow s^p;d$ is a forgetting rule. An extended spiking neural P system with exhaustive use of rules~\cite{Ionescu2007B} applies its rules as follows. If a neuron $\sigma_i$ contains $k$ spikes and the rule $E/s^b\rightarrow s^p;d$ is applicable, then the neuron $\sigma_i$ sends out $gp$ spikes after $d$ timesteps leaving $u$ spikes in $\sigma_i$, where $k=bg+u$, $u<b$ and $k,g,u\in\Nset$. Thus, a synapse in a spiking neural P system with exhaustive use of rules may transmit an arbitrary number of spikes in a single timestep. In the sequel we allow the input neuron of a system with exhaustive use of rules to receive an arbitrary number of spikes in a single timestep. This is a generalisation on the input allowed by Ionescu et al.~\cite{Ionescu2007B}. We discuss why we think this generalisation is natural for this model at the end of the paper.

In earlier work~\cite{Paun2007}, Korec's notion of strong universality was adopted for small SN P systems. Analogously, some small SN P systems could be described as what Korec refers to as weak universality. However, as we noted in other work~\cite{Neary2009}, it could be considered that Korec's notion of strong universality is somewhat arbitrary and we also pointed out some inconsistency in his notion of weak universality. Hence, in this work we rely on time/space complexity analysis to compare the encodings used by the small SN P system in Table~\ref{tab:Small_SNP}.

In the sequel each spike in a spiking neural P system represents a single unit of space. The maximum number of spikes in a spiking neural P system at any given timestep during a computation is the space used by the system.

\section{Counter machines}\label{sec: counter machines} 
The definition we give for counter machine is similar to that of Fischer et al.~\cite{Fischer1968}.
\begin{definition}[Counter machine]\label{def:counter machine}\\
A counter machine is a tuple $C=(z,c_{m},Q,q_0,q_h,\Sigma,f)$, where $z$ gives the number of counters, $c_{m}$ is the output counter, $Q=\{q_0,q_1,\cdots,q_h\}$ is the set of states, $q_0,q_h\in Q$ are the initial and halt states respectively, $\Sigma$ is the input alphabet and $f$ is the transition function 
\begin{equation*}
f:(\Sigma\times Q \times g(i))\rightarrow (\{Y,N\}\times Q\times\{INC,DEC,NULL\}) 
\end{equation*}
 where $g(i)$ is a binary valued function and $0\leqslant i\leqslant z$, $Y$ and $N$ control the movement of the input read head, and $INC$, $DEC$, and $NULL$ indicate the operation to carry out on counter~$c_i$. 
\end{definition}
Each counter $c_i$ stores a natural number value $x$. If $x>0$ then $g(i)$ is true and if $x=0$ then $g(i)$ is false. The input to the counter machine is read in from an input tape with alphabet $\Sigma$. The movement of the scanning head on the input tape is one-way so each input symbol is read only once. When a computation begins the scanning head is over the leftmost symbol $\alpha$ of the input word $\alpha w\in\Sigma^\ast$ and the counter machine is in state $q_0$. We give three examples below to explain the operation of the transition function $f$. 
\begin{itemize}
\item $f(\alpha, q_j, g(i))= (Y, q_k, INC(h))$ move the read head right on the input tape to read the next input symbol, change to state $q_k$ and increment the value $x$ stored in counter $c_i$ by 1.
\item $f(\alpha, q_j, g(i))= (N, q_k, DEC(h))$ do not move the read head, change to state $q_k$ and decrement the value $x$ stored in counter $c_i$ by 1. Note that $g(i)$ must evaluate to true for this rule to execute.
\item $f(\alpha, q_j, g(i))= (N, q_k, NULL)$ do not move the read head and change to state $q_k$.
\end{itemize}
A single application of $f$ is a timestep. Thus in a single timestep only one counter may be incremented or decremented by 1.

Our definition for counter machine, given above, is more restricted than the definition given by Fischer~\cite{Fischer1968}. In Fischer's definition $INC$ and $DEC$ may be applied to every counter in the machine in a single timestep. Clearly the more general counter machines of Fischer simulate our machines with no extra space or time overheads.  Fischer has shown that counter machines are exponentially slow in terms of computation time as the following theorem illustrates.

\begin{theorem}[Fischer~\cite{Fischer1968}]\label{thm:counter machines exp time}
There is a language $L$, real-time recognizable by a one-tape TM, which is not recognizable by any $k$-CM in time less than $T(n)=2^{\frac{n}{2k}}$.
\end{theorem}

In Theorem~\ref{thm:counter machines exp time} a one-tape TM is an offline Turing machine with a single read only input tape and a single work tape, a $k$-CM is a counter machine with $k$ counters, $n$ is the input length and real-time recognizable means recognizable in $n$ timesteps. For his proof Fischer noted that the language $L=\{waw^r\mid w\in\{0,1\}^\ast\}$, where $w^r$ is $w$ reversed, is recognisable in $n$ timesteps on a one-tape offline Turing machine. He then noted, that time of $2^{\frac{n}{2k}}$ is required to process input words of length $n$ due to the unary data storage used by the counters of the $k$-CM. Note that Theorem~\ref{thm:counter machines exp time} also holds for non-deterministic counter machines as they use the same unary storage method.

\section{Non-deterministic counter machines simulate spiking neural P systems in linear time}\label{sec:Counter machines simulate P systems in linear time}
\begin{theorem}\label{thm:counter machines simulate SNPs in linear time} 
Let $\Pi$ be a spiking neural P system with $m$ neurons that completes its computation in time $T$ and space $S$. Then there is a non-deterministic counter machine $C_\Pi$ that simulates the operation of $\Pi$ in time $O(T(x_r)^2m+Tm^2)$ and space $O(S)$ where $x_r$ is a constant dependant on the rules of $\Pi$.
\end{theorem}

\subsubsection{Proof idea}
Before we give the proof of Theorem~\ref{thm:counter machines simulate SNPs in linear time} we give the main idea behind the proof. Each neuron $\sigma_i$ from the spiking neural P system $\Pi$ is simulated by a counter $c_i$ from the counter machine $C_\Pi$. If a neuron $\sigma_i$ contains $y$ spikes, then the counter will have value $y$. A single synchronous update of all the neurons at a given timestep $t$ is simulated as follows. If the number of spikes in a neuron $\sigma_i$ is deceasing by $b$ spikes in-order to execute a rule, then the value $y$ stored in the simulated neuron $c_i$ is decremented $b$ times using $DEC(i)$ to give $y-b$. This process is repeated for each neuron that executes a rule at time $t$. If neuron $\sigma_i$ fires at time $t$ and has synapses to neurons $\{\sigma_{i_1},\ldots\sigma_{i_v}\}$ then for each open neuron $\sigma_{i_j}$ in $\{\sigma_{i_1},\ldots\sigma_{i_v}\}$ at time $t$ we increment the simulated neuron $c_{i_j}$ using $INC(i_j)$. This process is repeated until all firing neurons have been simulated. This simulation of the synchronous update of $\Pi$ at time $t$ is completed by $C_\Pi$ in constant time. Thus we get the linear time bound given in Theorem~\ref{thm:counter machines simulate SNPs in linear time}.

\begin{proof}
Let $\Pi =(O,\sigma_1,\sigma_2,\cdots ,\sigma_m,syn,in,out)$ be a spiking neural P system where $in=\sigma_1$ and $out=\sigma_2$. We explain the operation of a non-deterministic counter machine $C_{\Pi}$ that simulates the operation of $\Pi$ in time $O(T(x_r)^2m+Tm^2)$ and space $O(S)$. 

There are $m+1$ counters $c_1,c_2,c_3,\cdots ,c_m,c_{m+1}$ in $C_{\Pi}$. Each counter $c_i$ emulates the activity of a neuron $\sigma_i$. If $\sigma_i$ contains $y$ spikes then counter $c_i$ will store the value $y$. The states of the counter machine are used to control which neural rules are simulated in each counter and also to synchronise the operations of the simulated neurons (counters). 

\subsubsection{Input encoding}
It is sufficient for $C_{\Pi}$ to have a binary input tape. The value of the binary word $w\in\{1,0\}^{\ast}$ that is placed on the terminal to be read into $C_{\Pi}$ is identical to the binary sequence read in from the environment by the input neuron $\sigma_i$. A single symbol is read from the terminal at each simulated timestep. The counter $c_1$ (the simulated input neuron) is incremented only on timesteps when a 1 (a simulated spike) is read. As such at each simulated timestep $t$, a simulated spike is received by $c_1$ if and only if a spike is received by the input neuron $\sigma_1$. At the start of the computation, before the input is read in, each counter simulating $\sigma_i$ is incremented $n_i$ times to simulated the $n_i$ spikes in each neuron given by 2(a) of Definition~\ref{def:Spiking neural P-systems}. This takes a constant amount of time.

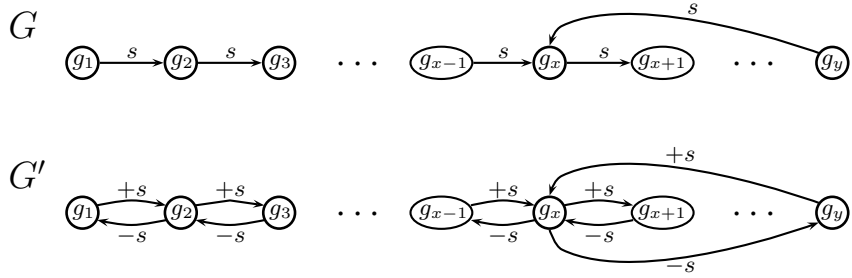
\begin{figure}[t]
\begin{center}
\psset{unit=2.4ex}
\begin{pspicture}(0,0)(30,4) %

\put(-2,2){\Large $G$}

\pscircle[linewidth=1pt](1,1){.7}
\put(.6,.9){$g_1$}
\psline{->}(1.7,1)(4.3,1)
\put(2.8,1.2){$s$}

\pscircle[linewidth=1pt](5,1){.7}
\put(4.6,.9){$g_2$}
\psline{->}(5.7,1)(8.3,1)
\put(6.8,1.2){$s$}

\pscircle[linewidth=1pt](9,1){.7}
\put(8.6,.9){$g_3$}
\put(11.3,.6){\Large $\cdots$}

\psellipse(15.6,1)(1.3,.7)
\put(14.7,.9){$g_{x-1}$}
\psline{->}(16.9,1)(19.3,1)
\put(17.9,1.2){$s$}

\pscircle[linewidth=1pt](20,1){.7}
\put(19.55,.9){$g_x$}
\psline{->}(20.7,1)(23.3,1)
\put(22,1.2){$s$}

\psellipse(24.6,1)(1.3,.7)
\put(23.6,.9){$g_{x+1}$}
\put(27.4,.6){\Large $\cdots$}

\pscircle[linewidth=1pt](31.5,1){.7}
\put(31.05,.9){$g_y$}
\pscurve{->}(30.95,1.4)(20.5,2.5)(20,1.7)
\put(25.6,3){$s$}
\end{pspicture}
\begin{pspicture}(0,-1.5)(30,6) %

\put(-2,2){\Large $G'$}

\pscircle[linewidth=1pt](1,1){.7}
\put(.6,.9){$g_1$}
\pscurve{->}(1.6,1.3)(3,1.5)(4.4,1.3)
\put(2.4,1.7){$+s$}
\pscurve{->}(4.4,.7)(3,.5)(1.6,0.7)
\put(2.4,-0.2){$-s$}

\pscircle[linewidth=1pt](5,1){.7}
\put(4.6,.9){$g_2$}
\psline{->}(5.6,1.3)(7,1.5)(8.4,1.3)
\put(6.4,1.7){$+s$}
\pscurve{->}(8.4,.7)(7,.5)(5.6,.7)
\put(6.4,-0.2){$-s$}

\pscircle[linewidth=1pt](9,1){.7}
\put(8.6,.9){$g_3$}
\put(11.3,.6){\Large $\cdots$}

\psellipse(15.6,1)(1.3,.7)
\put(14.7,.9){$g_{x-1}$}
\pscurve{->}(16.8,1.3)(18.1,1.5)(19.4,1.3)
\put(17.5,1.7){$+s$}
\pscurve{->}(19.4,0.7)(18.1,0.5)(16.8,0.7)
\put(17.5,-0.2){$-s$}

\pscircle[linewidth=1pt](20,1){.7}
\put(19.55,.9){$g_x$}
\pscurve{->}(20.6,1.3)(22,1.5)(23.4,1.3)
\put(21.4,1.7){$+s$}
\pscurve{->}(23.4,.7)(22,0.5)(20.6,.7)
\put(21.4,-0.2){$-s$}

\psellipse(24.6,1)(1.3,.7)
\put(23.6,.9){$g_{x+1}$}
\put(27.4,.6){\Large $\cdots$}
\pscircle[linewidth=1pt](31.5,1){.7}

\put(31.05,.9){$g_y$}
\pscurve{->}(30.95,1.4)(20.5,2.5)(20,1.7)
\put(24.7,3.1){$+s$}
\pscurve{->}(20,0.3)(20.5,-.5)(30.95,0.6)
\put(24.7,-1.4){$-s$}
\end{pspicture}
\end{center}
\caption{Finite state machine $G$ decides if a particular rule is applicable in a neuron given the number of spikes in the neuron \emph{at a given time} in the computation. Each $s$ represents a spike in the neuron. Machine $G'$ keeps track of the movement of spikes into and out of the neuron and decides whither or not a particular rule is applicable \emph{at each timestep} in the computation. $+s$ represents a single spike entering the neuron and $-s$ represents a single spike exiting the neuron.}\label{fig:finite_state_machine_G}
\end{figure}

\subsubsection{Storing neural rules in the counter machine states}
Recall from Definition~\ref{def:Spiking neural P-systems} that the applicability of a rule in a neuron is dependant on a regular expression over a unary alphabet. Let $r=E/s^{b}\rightarrow s;d$ be a rule in neuron $\sigma_i$. Then there is a finite state machine $G$ that accepts language $L(E)$ and thus decides if the number of spikes in $\sigma_i$ permits the application of $r$ in $\sigma_i$ at a given time in the computation. $G$ is given in Figure~\ref{fig:finite_state_machine_G}. If $g_j$ is an accept state in $G$ then $j>b$. This ensures that there is enough spikes to execute $r$. We also place the restriction on $G$ that $x>b$. During a computation we may use $G$ to decide if $r$ is applicable in $\sigma_i$ by passing an $s$ to $G$ each time a spike enters $\sigma_i$. However, $G$ may not give the correct result if spikes leave the neuron as it does not record spikes leaving $\sigma_i$. Thus using $G$ we may construct a second machine $G'$ such that $G'$ records the movement of spikes going into and out of the neuron. $G'$ is construct as follows; $G'$ has all the same states (including accept states) and transitions as $G$ along with an extra set of transitions that record spikes leaving the neuron. This extra set of transitions are given as follows for each transition on $s$ from a state $g_i$ to a state $g_j$ in $G$ there is a new transition on $-s$ going from state $g_i$ to $g_j$ in $G'$ that records the removal of a spike from $G'$.
By recording the dynamic movement of spikes, $G'$ is able to decide if the number of spikes in $\sigma_i$ permits the application of $r$ in $\sigma_i$ at each timestep during the computation. $G'$ is also given in Figure~\ref{fig:finite_state_machine_G}. Note that forgetting rules $s^{e}\rightarrow \lambda;0$ are dependant on simpler regular expressions thus we will not give a machine $G'$ for forgetting rules here.

Let neuron $\sigma_i$ have the greatest number $l$ of rules of any neuron in $\Pi$. Thus the applicability of rules $r_1,r_2,\cdots,r_l$ in $\sigma_i$ is decided by the automata $G'_1,G'_2,\cdots,G'_l$. We record if a rule may be simulated in a neuron at any given timestep during the computation by recording the current state of its $G'$ automaton (Figure~\ref{fig:finite_state_machine_G}) in the states of the counter machine. There are $m$ neuron in~$\Pi$. Thus each state in our counter machine remembers the current states of at most $ml$ different $G'$ automata in order to determine which rules are applicable in each neuron at a given time. 

Recall that in each rule of the form $r=E/s^b\rightarrow s;d$ that $d$ specifies the number of timestep between the removal of $b$ spikes from the neuron and the spiking of the neuron. The number of timesteps $<d$ remaining until a neuron will spike is recorded in the states of the $C_\Pi$. Each state in our counter machine remembers at most $m$ different values $<d$.

\subsubsection{Algorithm overview}
Next we explain the operation of $C_\Pi$ by explaining how it simulates the synchronous update of all neurons in $\Pi$ at an arbitrary timestep~$t$. The algorithm has 3 stages. A single iteration of Stage 1 identifies which applicable rule to simulate in a simulated open neuron. Then the correct number $y$ of simulated spikes are removed by decrementing the counter $y$ times ($y=b$ or $y=e$ in 2b of Definition~\ref{def:Spiking neural P-systems}). Stage 1 is iterated until all simulated open neurons have had the correct number of simulated spikes removed. A single iteration of Stage 2 identifies all the synapses leaving a firing neuron and increments every counter that simulates an open neuron at the end of one of these synapses. Stage 2 is iterated until all firing neurons have been simulated by incrementing the appropriate counters. Stage 3 synchronises each neuron with the global clock and increments the output counter if necessary. If the entire word $w$ has not been read from the input tape the next symbol is read.

\subsubsection{Stage 1. Identify rules to be simulated and remove spikes from neurons}
Recall that $d=0$ indicates a neuron is open and the value of $d$ in each neuron is recorded in the states of the counter machine. Thus our algorithm begins by determining which rule to simulate in counter $c_{i_1}$ where ${i_1}=min\{i\,|\,d=0\;for\,\sigma_i\}$ and the current state of the counter machine encodes an accept state for one or more of the $G'$ automata for the rules in $\sigma_{i_1}$ at time $t$. If there is more than one rule applicable the counter machine non-deterministically chooses which rule to simulate. Let $r=E/s^{b}\rightarrow s;d$ be the rule that is to be simulated. Using the $DEC(i_1)$ instruction, counter $c_{i_1}$ is decremented $b$ times. With each decrement of $c_{i_1}$ the new current state of each automaton $G'_1,G'_2,\cdots,G'_l$ is recorded in the counter machine's current state. After $b$ decrements of $c_i$ the simulation of the removal of $b$ spikes from neuron $\sigma_{i_1}$ is complete. Note that the value of $d$ from rule $r$ is recorded in the counter machine state.

There is a case not covered by the above paragraph. To see this note that in $G'$ in Figure~\ref{fig:finite_state_machine_G} there is a single non-deterministic choice to be made. This choice is at state $g_x$ if a spike is being removed ($-s$). Thus, if one of the automata is in such a state $g_x$ our counter machine resolves this be decrementing the counter $x$ times using the $DEC$ instruction. If $c_{i_1}=0$ after the counter has been decremented $x$ times then the counter machine simulates state $g_{x-1}$ otherwise state $g_{y}$ is simulated. Immediately after this the counter is incremented $x-1$ times to restore it to the correct value.

When the simulation of the removal of $b$ spikes from neuron $\sigma_{i_1}$ is complete, the above process is repeated with counter $c_{i_2}$ where ${i_2}=min\{i\,|\,{i_2}>{i_1},d=0\;for\,\sigma_i\}$ and the current state of the counter machine encodes an accept state for one or more of the $G'$ automata for the rules in $\sigma_{i_2}$ at time $t$. This process is iterated until every simulated open neuron with an applicable rule at time $t$ has had the correct number of simulated spikes removed.

\subsubsection{Stage 2. Simulate spikes}
This stage of the algorithm begins by simulating spikes traveling along synapses of the form $(i_1,j)$ where $i_1=min\{i\,|\,d=1\;for\,\sigma_i\}$ (if $d=1$ the neuron is firing). Let $\{(i_1,j_1),(i_1,j_2),\cdots,(i_1,j_k)\}$ be the set of synapses leaving $\sigma_i$ where $j_u<j_{u+1}$ and $d\leqslant 1$ in $\sigma_{j_u}$ at time $t$ (if $d\leqslant 1$ the neuron is open and may receive spikes). Then the following sequence of instructions are executed INC$(j_1)$, INC$(j_2)$, $\cdots$, INC$(j_k)$, thus incrementing any counter (simulated neuron) that receives a simulated spike.

The above process is repeated for synapses of the form $(i_2,j)$ where $i_2=min\{i\,|\,i_2>i_1,d=1\;for\,\sigma_i\}$. This process is iterated until every simulated neuron $c_i$ that is open has been incremented once for each spike $\sigma_i$ receives at time~$t$.

\subsubsection{Stage 3. Reading input, decrementing $d$, updating output counter and halting}
If the entire word $w$ has not been read from the input tape then the next symbol is read. If this is the case and the symbol read is a 1 then counter $c_1$ is incremented thus simulating a spike being read in by the input neuron. In this stage the state of the counter machine changes to record the fact that each $k\leqslant d$ that records the number of timesteps until a currently closed neuron will fire is decremented to $k-1$. If the counter $c_m$, which simulates the output neuron, has spiked only once prior to the simulation of timestep $t+1$ then this stage will also increment output counter $c_{m+1}$. If during the simulation of timestep $t$ counter $c_m$ has simulated a spike for the second time in the computation, then the counter machine enters the halt state. When the halt state is entered the number stored in counter $c_{m+1}$ is equal to the unary output that is given by time between the first two spikes in $\sigma_{m}$.

\subsubsection{Space analysis}
The input word on the binary tape of $C_\Pi$ is identical to the length of the binary sequence read in by the input neuron of $\Pi$. Counters $c_1$ to $c_m$ uses the same space as neurons $\sigma_1$ to $\sigma_m$. Counter $c_{m+1}$ uses the same amount of space as the unary output of the computation of $\Pi$. Thus $C_\Pi$ simulates $\Pi$ in space of $O(S)$. 
\subsubsection{Time analysis}
The simulation involves 3 stages. Recall that $x>b$. Let $x_r$ be the maximum value for $x$ of any $G'$ automaton thus $x_r$ is greater than the maximum number of spikes deleted in a neuron.

Stage 1. In order to simulate the deletion of a single spike in the worst case the counter will have to be decremented $x_r$ times and incremented $x_r-1$ times as in the special case. This is repeated a maximum of $b<x_r$ times (where $b$ is the number of spikes removed). Thus a single iteration of Stage 1 take $O({x_r}^2)$ time. Stage 1 is iterated a maximum of $m$ times per simulated timestep giving $O({x_r}^2m)$ time.
 
Stage 2. The maximum number of synapses leaving a neuron $i$ is $m$. A single spike traveling along a neuron is simulated in one step. Stage 2 is iterated a maximum of $m$ times per simulated timestep giving $O(m^2)$ time. 

Stage 3. Takes a small constant number of steps.

Thus a single timestep of $\Pi$ is simulated by $C_\Pi$ in $O((x_r)^2m+m^2)$ time and $T$ timesteps of $\Pi$ are simulated in linear time $O(T(x_r)^2m+Tm^2)$ by $C_\Pi$.
\qed
\end{proof}

The following is an immediate corollary of Theorems~\ref{thm:counter machines exp time} and~\ref{thm:counter machines simulate SNPs in linear time}.

\begin{corollary}
There exist no universal spiking neural P system that simulates Turing machines with less than exponential time and space overheads. 
\end{corollary}

\section{A universal spiking neural P system that is both small and time efficient}\label{sec:small time efficient SNP system}
In this section we construct a universal spiking neural P system that applies exhaustive use of rules, has only 10 neurons, and simulates any Turing machine in linear time.

\begin{theorem}\label{thm:poly time SNP system}
Let $M$ be a single tape Turing machine with $|A|$ symbols and $|Q|$ states that runs in time $T$. Then there is a universal spiking neural P system $\Pi_{M}$ with exhaustive use of rules that simulates the computation of $M$ in time $O(|A||Q|T)$ and space $O({[2^{\log_2\lceil 2|Q||A|+2|A|\rceil}}]^{T})$ and has only 10 neurons.
\end{theorem}

If the reader would like to get a quick idea of how our spiking neural P system with 10 neurons operates they should skip to the algorithm overview in Subsection~\ref{sec:Algorithm overview} of the proof.

\begin{proof}
We give a spiking neural P system $\Pi_{M}$ that simulates an arbitrary Turing machine $M$ in linear time and exponential space. $\Pi_{M}$ is given by Figure~\ref{fig:universal SNP} and Tables~\ref{tab:neurons of universal SNP I} and~\ref{tab:neurons of Extended SNP II}. The algorithm for $\Pi_{M}$ is deterministic and is mainly concerned with the simulation of an arbitrary transition rule. Without loss of generality we insist that $M$ always finishes its computation with the tape head at the leftmost end of the tape contents. Let $M$ be any single tape Turing machine with symbols $\alpha_1,\alpha_2,\ldots,\alpha_{|A|}$ and states $q_1,q_2,\ldots q_{|Q|}$, blank symbol $\alpha_1$, and halt state $q_{|Q|}$.

\subsection{Encoding a configuration of Turing machine $M$}\label{sec:Encoding a config of M} 
Each configuration of $M$ is encoded as three natural numbers using a well known technique. A configuration of $M$ is given by the following equation
\begin{equation}\label{eq:configuration}
C_k=\,\pmb{q_r,}\;\cdots \alpha_1\alpha_1\alpha_1\, a_{-x}\cdots a_{-3}a_{-2}a_{-1}\underline{a_{0}}a_{1}a_{2}a_{3} \cdots a_{y}\, \alpha_1\alpha_1\alpha_1 \cdots
\end{equation}
where $q_r$ is the current state, each $a_i$ is a tape cell of $M$ and the tape head of $M$, given by an underline, is over $a_0$. Also, tape cells $a_{-x}$ and $a_{y}$ both contain $\alpha_1$, and the cells between $a_{-x}$ and $a_{y}$ include all of the cells on $M$'s tape that have either been visited by the tape head prior to configuration $C_k$ or contain part of the input to $M$.

In the sequel the encoding of object $p$ is given by $\langle p \rangle$.
The tape symbols $\alpha_1,\alpha_2,\ldots,\alpha_{|A|}$ of $M$ are encoded as $\langle \alpha_1\rangle=1,\langle \alpha_2\rangle=3,\ldots, \langle \alpha_{|A|}\rangle=2|A|-1$, respectively, and the states $q_1,q_2,\ldots, q_{|Q|}$ are encoded as $\langle q_1\rangle=2A,\langle q_2\rangle=4A,\ldots, \langle q_{|Q|}\rangle=2|Q|A$, respectively. The contents of each tape cell $a_i$ in configuration $C_k$ is encoded as $\langle a_i \rangle = \langle \alpha \rangle$ where $\alpha$ is a tape symbol of $M$. The tape contents in Equation~\eqref{eq:configuration} to the left and right of the tape head are respectively encoded as the numbers $X=\underset{i=1}{\overset{x}{\sum }}z^i\langle a_{-i} \rangle$ and $Y=\underset{j=1}{\overset{y}{\sum }}z^j\langle a_j\rangle$ where $z=2^v$ and $v=\lceil\log_2(2|Q||A|+2|A|)\rceil$. Thus the entire configuration $C_k$ is encoded as three natural numbers via the equation 
\begin{equation}\label{eq:encoded configuration}
\langle C_k\rangle\,=\,\left(X,\; Y,\; \langle q_r\rangle+\langle\alpha_i\rangle\right)
\end{equation}
where $\langle C_k\rangle$ is the encoding of $C_k$ from Equation~\eqref{eq:configuration} and $\alpha_i$ is the symbol being read by the tape head in cell $a_0$. 

A transition rule $q_r,\alpha_i,\alpha_j,D,q_u$ of $M$ is executed on $C_k$ as follows. If the current state is $q_r$ and the tape head is reading the symbol $\alpha_i$ in cell $a_0$, $\alpha_j$ the write symbol is printed to cell $a_0$, the tape head moves one cell to the left to $a_{-1}$ if $D=L$ or one cell to the right to $a_1$ if $D=R$, and $q_u$ becomes the new current state. A simulation of transition rule $q_r,\alpha_i,\alpha_j,D,q_u$ on the encoded configuration $\langle C_k\rangle $ from Equation~\eqref{eq:encoded configuration} is given by the equation
\begin{equation}\label{eq:transition rule simulation}
\langle C_{k+1}\rangle= 
\begin{cases}
\left(\frac{X}{z}-(\frac{X}{z}\mod z),\;zY+z\langle\alpha_j\rangle,\;\langle q_u\rangle+(\frac{X}{z}\mod z)\right)&
\\
\left(zX+z\langle \alpha_j\rangle,\;\frac{Y}{z}-(\frac{Y}{z}\mod z),\;\langle q_u\rangle+(\frac{Y}{z}\mod z)\right)&\\ 
\end{cases}
\end{equation}
where configuration $C_{k+1}$ results from executing a single transition rule on configuration $C_{k}$, and $(b\mod c)=d$ where $d<c$, $b=ec+d$ and $b,c,d,e\in\Nset$. In Equation~\eqref{eq:transition rule simulation} the top case is simulating a left move transition rule and the bottom case is simulating a right move transition rule. In the top case, following the left move, the sequence to the right of the tape head is longer by 1 tape cell, as cell $a_0$ is added to the right sequence. Cell $a_0$ is overwritten with the write symbol $\alpha_j$ and thus we compute $zY+z\langle \alpha_j\rangle$ to simulate cell $a_0$ becoming part of the right sequence. Also, in the top case the sequence to the left of the tape head is getting shorter by 1 tape cell thus we compute $\frac{X}{z}-(\frac{X}{z}\mod z)$. The rightmost cell of the left sequence $a_{-1}$ is the \emph{new} tape head location and the tape symbol it contains is encoded as $(\frac{X}{z}\mod z)$. Thus the value $(\frac{X}{z}\mod z)$ is added to the new encoded current state $\langle q_u\rangle$. For the bottom case, a right move, the sequence to the right gets shorter which is simulated by $\frac{Y}{z}-(\frac{Y}{z}\mod z)$ and the sequence to the left gets longer which is simulated by $zX+z\langle\alpha_j\rangle$. The leftmost cell of the right sequence $a_1$ is the new tape head location and the tape symbol it contains is encoded as $(\frac{Y}{z}\mod z)$.

\subsection{Input to $\Pi_{M}$} 
Here we give an explanation of how the input is read into $\Pi_{M}$. We also give a rough outline of how the input to $\Pi_{M}$ is encoded in linear time.

A configuration $C_k$ given by Equation~\eqref{eq:encoded configuration} is read into $\Pi_{M}$ as follows. All the neurons of the system initially have no spikes with the exception of $\sigma_{10}$ which has 31 spikes. The input neuron $\sigma_5$ receives $X+2$ spikes at the first timestep $t_1$, $Y$ spikes at time $t_2$, and $\langle q_r\rangle+\langle\alpha_i\rangle$ spikes at time $t_4$. 
We explain how the system is initialised to encode an initial configuration of $M$ by giving the number of spikes in each neuron and the rule that is to be applied in each neuron at time~$t$. Thus at time $t_1$ we have
\begin{xalignat*}{2}
t_{1}:\;	&\sigma_5=X+2,	&s^{2}(s^{z})^\ast /s\rightarrow s;1, \\
		&\sigma_{10}=31,&s^{31}/ s^{16}\rightarrow \lambda;0.
\end{xalignat*}
where on the left $\sigma_j=k$ gives the number $k$ of spikes in neuron $\sigma_j$ at time $t_i$ and on the right is the next rule that is to be applied at time $t_i$ if there is an applicable rule at that time. Thus from Figure~\ref{fig:universal SNP} when we apply the rule $s^{2}(s^{z})^\ast /s\rightarrow s;1$ in neuron $\sigma_5$ and the rule $s^{31}/ s^{16}\rightarrow \lambda;0$ in neuron $\sigma_{10}$ at time $t_1$ we get
\begin{xalignat*}{2}
t_{2}:\;	&\sigma_4=X+2,	&s^{2}(s^{z})^\ast /s^{z}\rightarrow s^{z};2,\\
		&\sigma_5=Y,	&s^{2z}(s^{z})^\ast /s\rightarrow s;1, \\
		&\sigma_6,\sigma_7,\sigma_8,\sigma_9=X+2,	&s^{2}(s^{z})^\ast /s\rightarrow \lambda;0,\\
		&\sigma_{10}=15,&s^{15}/ s^8\rightarrow \lambda;0.\\
\\
t_{3}:\;	&\sigma_4=X+2,	&s^{2}(s^{z})^\ast /s^{z}\rightarrow s^{z};1,\\
		&\sigma_6=Y,	&(s^{z})^\ast /s\rightarrow s;1, \\
		&\sigma_7,\sigma_8,\sigma_9=Y, &(s^{z})^\ast /s\rightarrow \lambda;0, \\
		&\sigma_{10}=7,	&s^7/ s^4\rightarrow \lambda;0.\\
\\
t_{4}:\;	&\sigma_1=X,	&\\
		&\sigma_2=Y,	& \\
		&\sigma_4=2,	&s^{2} /s^2\rightarrow \lambda;0,\\
		&\sigma_5=\langle q_r\rangle+\langle \alpha_i\rangle,	&(s^{z})^\ast s^{\langle q_r\rangle+\langle \alpha_i\rangle} /s\rightarrow s;1,\\
		&\sigma_{10}=3,	&s^3/ s^2\rightarrow \lambda;0.
\end{xalignat*}
\begin{xalignat*}{2}
t_{5}:\;	&\sigma_1=X,	&\\
		&\sigma_2=Y,	& \\
		&\sigma_4,\sigma_6=\langle q_r\rangle+\langle \alpha_i\rangle,	&\\
		&\sigma_7,\sigma_8,\sigma_9=\langle q_r\rangle+\langle \alpha_i\rangle, &s^{\langle q_r\rangle+\langle \alpha_i\rangle} /s\rightarrow \lambda;0, \\
		&\sigma_{10}=1,	&s/ s\rightarrow s;\log_2(z)+3.\\
\end{xalignat*}
Forgetting rules are applied to get rid of superfluous spikes (for example see neurons $\sigma_7,\sigma_8,$ and $\sigma_9$ at time $t_2$). Note that $\sigma_4$ is closed at time $t_2$ as there is a delay of 2 on the rule ($s^{2}(s^{z})^\ast /s^{z}\rightarrow s^{z};2$) to be executed in $\sigma_4$. This prevents the $Y$ spikes from entering neuron $\sigma_4$ when $\sigma_5$ fires at time $t_2$. At time $t_5$ the spiking neural P system has $X$ spikes in $\sigma_1$, $Y$ spikes in $\sigma_2$, and $\langle q_r\rangle+\langle \alpha_i\rangle$ spikes in $\sigma_4$ and $\sigma_6$. Thus at time $t_{5}$ the spiking neural P system encodes an initial configuration of $M$. 

In this paragraph we will show that given an initial configuration of $M$ it is encoded as input to our spiking neural P system in Figure~\ref{fig:universal SNP} in linear time. In order to do this we must compute the three numbers that give $\langle C_k \rangle$ from Equation~\ref{eq:encoded configuration} in linear time. The number $X$ is computed as follows: given a sequence $a_{-x}a_{-x+1}\ldots a_{-2}a_{-1}$ the sequence $w=\langle a_{-x}\rangle 0^{\log_2(z)-1}\langle a_{-x+1}\rangle 0^{\log_2(z)-1} \ldots \langle a_{-2}\rangle 0^{\log_2(z)-1} \langle a_{-1} \rangle 0^{\log_2(z)-1}2$ is easily computed in time that is linear in $x$. The spiking neural P system $\Pi_{input}$ in Figure~\ref{fig:input SNP} takes the sequence $w$ and converts it into the $X$ spikes that form part of the input to our system in Figure~\ref{fig:universal SNP}. We give a rough idea of how $\Pi_{input}$ operates (if the reader wishes to pursue a more detailed view the rules for $\Pi_{input}$ are to be found in Table~\ref{tab:rules for input SNP}). The input neuron of $\Pi_{input}$ receives the sequence $w$ as a sequence of spikes and no-spikes. On each timestep where $\langle a\rangle$ is read $\langle a\rangle$ spikes are passed to the input neuron $\sigma_1$, and on each timestep where 0 is read no spikes are passed to the input neuron. Thus at timestep $t_1$ neuron $\sigma_1$ receives $\langle a_{-x}\rangle$ spikes, and at timestep $t_2$ neurons $\sigma_2$, $\sigma_3$, and $\sigma_4$ receive $\langle a_{-x}\rangle$ spikes from $\sigma_1$. Following timestep $t_2$, the number of spikes in neurons $\sigma_2$, $\sigma_3$, and $\sigma_4$ double with each timestep. So at timestep $t_{\log_2(z)+1}$ the number of spikes in each of the neurons $\sigma_2$, $\sigma_3$, and $\sigma_4$ is $\frac{z}{2}\langle a_{-x}\rangle$. At timestep $t_{\log_2(z)+1}$ neurons $\sigma_2$, $\sigma_3$ and $\sigma_4$ also receive $\langle a_{-x+1}\rangle$ spikes from $\sigma_1$ giving a total of $z\langle a_{-x}\rangle+\langle a_{-x+1}\rangle$ spikes in each of these neurons at time $t_{\log_2(z)+2}$. Proceeding to time $t_{2\log_2(z)+2}$ neurons $\sigma_2$, $\sigma_3$ and $\sigma_4$
have $z^2\langle a_{-x}\rangle+z\langle a_{-x+1}\rangle+\langle a_{-x+2}\rangle$ spikes. This process continues until $X=\underset{i=1}{\overset{x}{\sum }}z^i\langle a_{-i} \rangle$ is computed. The end of the process is signaled when the rightmost number in the sequence is read. When this number (2) is read it allows the result to be passed to $\sigma_6$ via $\sigma_5$. Following this $\sigma_6$ sends $X$ spikes out of the system. Note that prior to this 2 being read only forgetting rules are executed in $\sigma_6$ thus preventing any spikes from being sent out of the system. $\Pi_{input}$ computes $X$ in time $x\log_2(z)+3$. Recall from Section~\ref{sec:Encoding a config of M} that the value of $z$ is dependant on the number of states and symbols in $M$ thus $X$ is computed in time that is linear in $x$. In a similar manner, the value $Y$ is computed by $\Pi_{input}$ in time linear in $y$. The number $\langle q_r\rangle+\langle \alpha_i\rangle$ is computed in constant time. Thus the input $\langle C_k \rangle$ for $\Pi_{M}$ is computed in linear time.

\begin{figure}[t]
         \begin{center}
                    \begin{tikzpicture}[>=stealth',shorten >=1pt,auto,node distance=1.5cm,thick,bend angle=45]

                        \tikzstyle{every state}=[ellipse,inner sep=0pt, node distance=2.3cm]
			\tikzstyle{state1}=[circle,draw=black!75, node distance=1.5cm]
			\tikzstyle{state2}=[circle,draw=black!75, node distance=1.3cm]
			\tikzstyle{input-output}=[draw=none,node distance=1cm]
			\tikzstyle{dummy}=[draw=none,node distance=1.4cm]

			\node[state]    	(sigma1)	[]	{};		\draw (sigma1)+(0,0) node {\large $\sigma_1$};
			\node[input-output]    	(input)		[above of=sigma1]	{};		\draw (input)+(0,.1) node {\large input};
			\node[state]    	(sigma2)	[below of=sigma1]	{$\quad\;\;$};	\draw (sigma2)+(0,0) node {\large $\sigma_2$};
			\node[state2]    	(sigma3)	[right of=sigma2]	{$\quad\;\;$};	\draw (sigma3)+(0,0) node {\large $\sigma_3$};
			\node[state2]    	(sigma4)	[right of=sigma3]	{$\quad\;\;$};	\draw (sigma4)+(0,0) node {\large $\sigma_{4}$};
			\node[state]    	(sigma5)	[below of=sigma2]	{};		\draw (sigma5)+(0,0) node {\large $\sigma_{5}$};
			\node[state]    	(sigma6)	[right of=sigma5]	{};		\draw (sigma6)+(0,0) node {\large $\sigma_{6}$};
			\node[input-output]    	(output)	[below of=sigma6]	{};		\draw (output)+(0,-.1) node {\large output};

			\path [->]	(input)	edge			node {}	(sigma1)
					(sigma1)edge			node {}	(sigma2)	
					(sigma1)edge			node {}	(sigma3)
					(sigma1)edge			node {}	(sigma4)
					(sigma1)edge[bend right,in=230]node {}	(sigma5)
					(sigma2)edge			node {}	(sigma5)
					(sigma4)edge			node {}	(sigma5)
					(sigma5)edge			node {}	(sigma6)
					(sigma6)edge			node {}	(output);

			\path [<->]	(sigma2)edge			node {}	(sigma3)
					(sigma2)edge[bend right,in=220]	node {}	(sigma4)
					(sigma3)edge			node {}	(sigma4);

		\end{tikzpicture}
	\end{center}
	\caption{Spiking neural P system $\Pi_{input}$. Each circle is a neuron and each arrow represents the direction spikes move along a synapse between a pair of neurons. The rules for $\Pi_{input}$ are to be found in Table~\ref{tab:rules for input SNP}.}\label{fig:input SNP}
\end{figure}
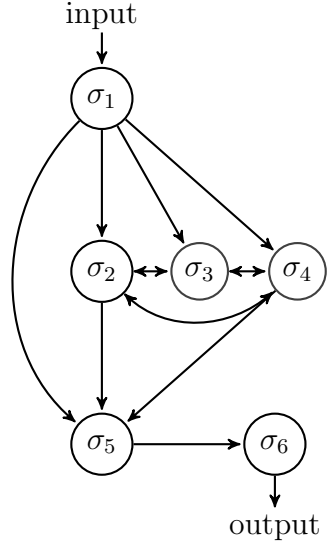

\subsection{Algorithm overview}\label{sec:Algorithm overview}
To help simplify the explanation, some of the rules given here differ slightly from those in the more detailed simulation that follows this overview. The numbers from Equation~\eqref{eq:encoded configuration}, encoding a Turing machine configuration, are stored in the neurons of our system as $X$, $Y$ and $\langle q_r\rangle+\langle\alpha_i\rangle$ spikes. Equation~\eqref{eq:transition rule simulation} is implemented in Figure~\ref{fig:universal SNP} to give a spiking neural P system $\Pi_{M}$ that simulates the transition rules of $M$. The two values $X$ and $Y$ are stored in neurons $\sigma_1$ and $\sigma_2$, respectively. If $X$ or $Y$ is to be multiplied the spikes that encode $X$ or $Y$ are sent down through the network of neurons from either $\sigma_1$ or $\sigma_2$ respectively, until they reach $\sigma_{10}$. Note in Figure~\ref{fig:universal SNP} that each neuron from $\sigma_7,\sigma_8$ and $\sigma_{9}$ has incoming synapses coming from the other two neurons in $\sigma_7,\sigma_8$ and $\sigma_{9}$. Thus if $\sigma_7,\sigma_8$ and $\sigma_{9}$ each contain $N$ spikes at time $t_k$, and they each fire sending $N$ spikes, then each of the neurons $\sigma_7,\sigma_8$ and $\sigma_{9}$ will contain $2N$ spikes at time $t_{k+1}$. Given $Y$ the value $zY=2^vY$ is computed as follows: First we calculate $2Y$ by firing $\sigma_7,\sigma_8$ and $\sigma_{9}$, then $4Y$ by firing $\sigma_7,\sigma_8$, and $\sigma_{9}$ again. After $v$ timesteps the value $zY$ is computed. $zX$ is computed using the same technique.

Now, we give the general idea of how the neurons compute $\frac{X}{z}-(\frac{X}{z}\mod z)$ and $(\frac{X}{z}\mod z)$ from Equation~\eqref{eq:transition rule simulation} (a slightly different strategy is used in the simulation). We begin with $X$ spikes in $\sigma_1$. The rule $(s^{z})^\ast / s^{z}\rightarrow s;1$ is applied in $\sigma_1$ sending $\frac{X}{z}$ spikes to $\sigma_4$. Following this $(s^{z})^\ast s^{(\frac{X}{z}\mod z)}/s^{z} \rightarrow s^{z};1$ is applied in $\sigma_4$ which sends $\frac{X}{z}-(\frac{X}{z}\mod z)$ to $\sigma_1$ leaving $(\frac{X}{z}\mod z)$ spikes in $\sigma_4$. The values $\frac{Y}{z}-(\frac{Y}{z}\mod z)$ and $(\frac{Y}{z}\mod z)$ are computed in a similar manner. 

Finally, using the encoded current state $\langle q_r\rangle$ and the encoded read symbol $\langle \alpha_i\rangle$ the values $z\langle \alpha_j\rangle$ and $\langle q_u\rangle$ from Equation~\eqref{eq:transition rule simulation} are computed. Using the technique outlined in the first paragraph of the algorithm overview the value $z(\langle q_r\rangle+\langle\alpha_i\rangle)$ is computed by sending $\langle q_r\rangle+\langle\alpha_i\rangle$ spikes from $\sigma_5$ to $\sigma_{10}$ in Figure~\ref{fig:universal SNP}. Then the rule $s^{z(\langle q_r\rangle+\langle\alpha_i\rangle)}/ s^{z(\langle q_r\rangle+\langle\alpha_i\rangle)-\langle q_u\rangle}\rightarrow s^{z\langle \alpha_j\rangle};1$ is applied in $\sigma_{10}$ which sends $z\langle \alpha_j\rangle$ spikes out to neurons $\sigma_4$ and $\sigma_6$. This rule uses $z(\langle q_r\rangle+\langle\alpha_i\rangle)-\langle q_u\rangle$ spikes thus leaving $\langle q_u\rangle$ spikes remaining in $\sigma_{10}$. This completes our sketch of how $\Pi_{M}$ in Figure~\ref{fig:universal SNP} computes the values in Equation~\eqref{eq:transition rule simulation} to simulate a transition rule. A more detailed simulation of a transition rule follows.

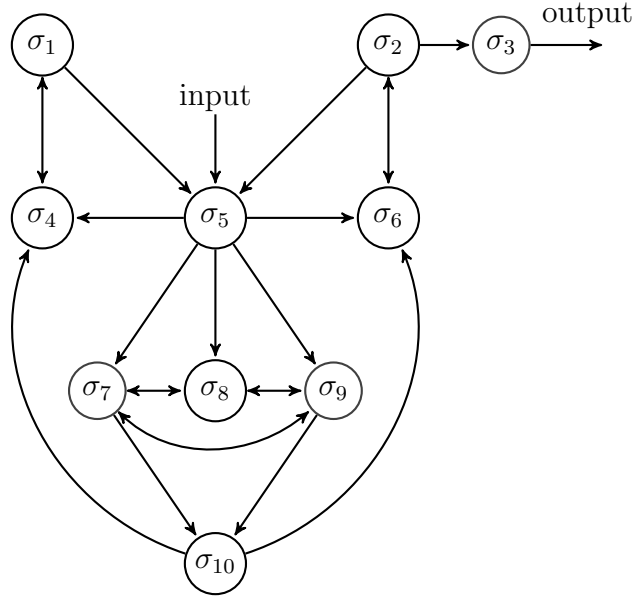
\begin{figure}[t]
         \begin{center}
                    \begin{tikzpicture}[>=stealth',shorten >=1pt,auto,node distance=1.5cm,thick,bend angle=45]

                        \tikzstyle{every state}=[ellipse,inner sep=0pt, node distance=2.3cm]
			\tikzstyle{state1}=[circle,draw=black!75, node distance=1.5cm]
			\tikzstyle{state2}=[circle,draw=black!75, node distance=1.57cm]
			\tikzstyle{input-output}=[draw=none,node distance=1.5cm]
			\tikzstyle{dummy}=[draw=none,node distance=1.4cm]

			\node[state]		(sigma1)				{};		\draw (sigma1)+(0,0) node {\large $\sigma_1$};
			\node[state]    	(sigma4)	[below of=sigma1]	{};		\draw (sigma4)+(0,0) node {\large $\sigma_4$};
			\node[state]    	(sigma5)	[right of=sigma4]	{};		\draw (sigma5)+(0,0) node {\large $\sigma_5$};
			\node[input-output]    	(input)		[above of=sigma5]	{};		\draw (input)+(0,.1) node {\large input};
			\node[state]    	(sigma6)	[right of=sigma5]	{};		\draw (sigma6)+(0,0) node {\large $\sigma_6$};
			\node[state]		(sigma2)	[above of=sigma6]	{};		\draw (sigma2)+(0,0) node {\large $\sigma_2$};
			\node[state1]		(sigma3)	[right of=sigma2]	{$\quad\;\;$};	\draw (sigma3)+(0,0) node {\large $\sigma_3$};
			\node[input-output]    	(output)	[right of=sigma3]	{};		\draw (output)+(-.35,.4) node {\large output};
			\node[state]    	(sigma8)	[below of=sigma5]	{};		\draw (sigma8)+(0,0) node {\large $\sigma_8$};
			\node[state2]    	(sigma7)	[left of=sigma8]	{$\quad\;\;$};	\draw (sigma7)+(0,0) node {\large $\sigma_7$};
			\node[state2]    	(sigma9)	[right of=sigma8]	{$\quad\;\;$};	\draw (sigma9)+(0,0) node {\large $\sigma_{9}$};
			\node[state]    	(sigma10)	[below of=sigma8]	{};		\draw (sigma10)+(0,0) node {\large $\sigma_{10}$};

			\path [->]	(input)	edge			node {}	(sigma5)
					(sigma1)edge			node {}	(sigma5)	
					(sigma2)edge			node {}	(sigma5)
					(sigma2)edge			node {}	(sigma3)
					(sigma3)edge			node {}	(output)
					(sigma5)edge			node {}	(sigma4)
					(sigma5)edge			node {}	(sigma6)
					(sigma5)edge			node {}	(sigma7)
					(sigma5)edge			node {}	(sigma8)
					(sigma5)edge			node {}	(sigma9)
					(sigma7)edge			node {}	(sigma10)
					(sigma9)edge			node {}	(sigma10)
					(sigma10)edge[bend right,in=230]node {}	(sigma6)
					(sigma10)edge[bend left,in=130]	node {}	(sigma4);

			\path [<->]	(sigma1)edge			node {}	(sigma4)	
					(sigma2)edge			node {}	(sigma6)
					(sigma7)edge			node {}	(sigma8)
					(sigma8)edge			node {}	(sigma9)
					(sigma7)edge[bend right,in=220]	node {}	(sigma9);

		\end{tikzpicture}
	\end{center}
	\caption{Universal spiking neural P system $\Pi_{M}$. Each circle is a neuron and each arrow represents the direction spikes move along a synapse between a pair of neurons. The rules for $\Pi_{M}$ are to be found in Tables~\ref{tab:neurons of universal SNP I} and~\ref{tab:neurons of Extended SNP II}.}\label{fig:universal SNP}
\end{figure}

\subsection{Simulation of $q_r,\alpha_i,\alpha_j,L,q_u$ (top case of Equation~\eqref{eq:transition rule simulation})}
The simulation of the transition rule begins at time $t_k$ with $X$ spikes in $\sigma_1$, $Y$ spikes in $\sigma_2$, $\langle q_r\rangle+\langle \alpha_i\rangle$ spikes in $\sigma_4$ and $\sigma_6$, and 1 spike in $\sigma_{10}$. As before we explain the simulation by giving the number of spikes in each neuron and the rule that is to be applied in each neuron at time~$t$. So at time $t_k$ we have
\begin{xalignat*}{2}
t_{k}:\;        & \sigma_1=X,                                     & \\
		&\sigma_2=Y,                                      & \\
		&\sigma_4,\sigma_6=\langle q_r\rangle+\langle\alpha_i\rangle,   &s^{\langle q_r\rangle+\langle\alpha_i\rangle}/ s\rightarrow s;1, \\
		&\sigma_{10}=1,    &s/ s\rightarrow s;\log_2(z)+3.
\end{xalignat*}
Thus from Figure~\ref{fig:universal SNP} when we apply the rule $s^{\langle q_r\rangle+\langle\alpha_i\rangle}/ s\rightarrow s;1$ in neurons $\sigma_4$ and $\sigma_6$ at time $t_k$ we get
\begin{xalignat*}{2}
t_{k+1}:\;	&\sigma_1=X+\langle q_r\rangle+\langle\alpha_i\rangle,	&s^{2z}(s^{z})^\ast s^{\langle q_r\rangle+\langle\alpha_i\rangle}/ s^{z}\rightarrow s;\log_2(z)+6, \\
		&\sigma_2=Y+\langle q_r\rangle+\langle\alpha_i\rangle,	&(s^{z})^\ast s^{\langle q_r\rangle+\langle\alpha_i\rangle}/ s\rightarrow s;1,\\
		&\sigma_{10}=1,    &s/ s\rightarrow s;\log_2(z)+2.
\end{xalignat*}
\begin{xalignat*}{2}
t_{k+2}:\;	&\sigma_1=X+\langle q_r\rangle+\langle\alpha_i\rangle,	&s^{2z}(s^{z})^\ast s^{\langle q_r\rangle+\langle\alpha_i\rangle}/ s^{z}\rightarrow s;\log_2(z)+5, \\
		&\sigma_3=Y+\langle q_r\rangle+\langle\alpha_i\rangle,\\
  &\qquad\qquad\qquad\textrm{if } \langle q_r\rangle=\langle q_{|Q|}\rangle &(s^{z})^\ast s^{\langle q_r\rangle+\langle\alpha_i\rangle}/ s^z\rightarrow s^z;1,\\
		&\qquad\qquad\qquad \textrm{if } \langle q_r\rangle\neq\langle q_{|Q|}\rangle &(s^{z})^\ast s^{\langle q_r\rangle+\langle\alpha_i\rangle}/ s\rightarrow \lambda;0,\\
		&\sigma_5=Y+\langle q_r\rangle+\langle\alpha_i\rangle,    &(s^{z})^\ast s^{\langle q_r\rangle+\langle\alpha_i\rangle}/ s\rightarrow s;1,\\
		&\sigma_6=Y+\langle q_r\rangle+\langle\alpha_i\rangle,    &s^{z}(s^{z})^\ast s^{\langle q_r\rangle+\langle\alpha_i\rangle}/ s\rightarrow \lambda;0,\\
		&\sigma_{10}=1,    &s/ s\rightarrow s;\log_2(z)+1.\\
\\ 
t_{k+3}:\;	&\sigma_1=X+\langle q_r\rangle+\langle\alpha_i\rangle,  &s^{2z}(s^{z})^\ast s^{\langle q_r\rangle+\langle\alpha_i\rangle}/ s^{z}\rightarrow s;\log_2(z)+4, \\
		&\sigma_4,\sigma_6=Y+\langle q_r\rangle+\langle\alpha_i\rangle,    &s^{z}(s^{z})^\ast s^{\langle q_r\rangle+\langle\alpha_i\rangle}/ s\rightarrow \lambda;0,\\
		&\sigma_7,\sigma_8,\sigma_{9}=Y+\langle q_r\rangle+\langle\alpha_i\rangle,    &s^{z}(s^{z})^\ast s^{\langle q_r\rangle+\langle\alpha_i\rangle}/ s\rightarrow s;1,\\
		&\sigma_{10}=1,    &s/ s\rightarrow s;\log_2(z).
\end{xalignat*}
In timestep $t_{k+2}$ above $\sigma_3$ the output neuron fires if and only if the encoded current state encodes the halt state $q_{|Q|}$. Recall that when $M$ halts the entire tape contents are to the right of the tape head, thus only $Y$ the encoding of the right sequence is sent out of the system. Thus the unary output is a number of spikes that encodes the tape contents of $M$.

Note that at timestep $t_{k+3}$ the neuron $\sigma_{7}$ receives $Y+\langle q_r\rangle+\langle\alpha_i\rangle$ spikes from each of the two neurons $\sigma_{8}$ and $\sigma_{9}$. Thus at time $t_{k+4}$ neuron $\sigma_{7}$ contains $2(Y+\langle q_r\rangle+\langle\alpha_i\rangle)$ spikes. In a similar manner $\sigma_{8}$ and $\sigma_{9}$ also receive $2(Y+\langle q_r\rangle+\langle\alpha_i\rangle)$ spikes at timestep $t_{k+3}$. The number of spikes in each of the neurons $\sigma_{7}$, $\sigma_{8}$ and $\sigma_{9}$ doubles at each timestep between $t_{k+3}$
and $t_{k+\log_2(z)+2}$. 
\begin{xalignat*}{2}
t_{k+4}:\;      &\sigma_1=X+\langle q_r\rangle+\langle\alpha_i\rangle,  &s^{2z}(s^{z})^\ast s^{\langle q_r\rangle+\langle\alpha_i\rangle}/ s^{z}\rightarrow s;\log_2(z)+3, \\
		&\sigma_7,\sigma_8,\sigma_{9}=2(Y+\langle q_r\rangle+\langle\alpha_i\rangle),    &s^{z}(s^{z})^\ast s^{2(\langle q_r\rangle+\langle\alpha_i\rangle)}/ s\rightarrow s;1,\\
		&\sigma_{10}=1,    &s/ s\rightarrow s;\log_2(z)-1.\\
\\
t_{k+5}:\;      &\sigma_1=X+\langle q_r\rangle+\langle\alpha_i\rangle,  &s^{2z}(s^{z})^\ast s^{\langle q_r\rangle+\langle\alpha_i\rangle}/ s^{z}\rightarrow s;\log_2(z)+2, \\
		&\sigma_7,\sigma_8,\sigma_{9}=4(Y+\langle q_r\rangle+\langle\alpha_i\rangle),    &s^{z}(s^{z})^\ast s^{4(\langle q_r\rangle+\langle\alpha_i\rangle)}/ s\rightarrow s;1,\\
		&\sigma_{10}=1,    &s/ s\rightarrow s;\log_2(z)-2.\\
\\
t_{k+6}:\;      &\sigma_1=X+\langle q_r\rangle+\langle\alpha_i\rangle,  &s^{2z}(s^{z})^\ast s^{\langle q_r\rangle+\langle\alpha_i\rangle}/ s^{z}\rightarrow s;\log_2(z)+1, \\
		&\sigma_7,\sigma_8,\sigma_{9}=8(Y+\langle q_r\rangle+\langle\alpha_i\rangle),    &s^{z}(s^{z})^\ast s^{8(\langle q_r\rangle+\langle\alpha_i\rangle)}/ s\rightarrow s;1,\\
		&\sigma_{10}=1,    &s/ s\rightarrow s;\log_2(z)-3.\\
\end{xalignat*}
The number of spikes in neurons $\sigma_{7}$, $\sigma_{8}$, and $\sigma_{9}$ continues to double until timestep $t_{k+\log_2(z)+2}$. When neurons $\sigma_{7}$ and $\sigma_{9}$ fire at timestep $t_{k+\log_2(z)+2}$ they send $\frac{z}{2}(Y+\langle q_r\rangle+\langle\alpha_i\rangle)$ spikes each to neuron $\sigma_{10}$ which has opened at time $t_{k+\log_2(z)+2}$ (for the first time in the transition rule simulation). Thus at time $t_{k+\log_2(z)+3}$ neuron $\sigma_{10}$ contains $z(Y+\langle q_r\rangle+\langle\alpha_i\rangle)$ spikes.
\begin{xalignat*}{2}
t_{k+\log_2(z)+2}:\;      &\sigma_1=X+\langle q_r\rangle+\langle\alpha_i\rangle,  &s^{2z}(s^{z})^\ast s^{\langle q_r\rangle+\langle\alpha_i\rangle}/ s^{z}\rightarrow s;5, \\
		&\sigma_7,\sigma_8,\sigma_{9}=\frac{z}{2}(Y+\langle q_r\rangle+\langle\alpha_i\rangle),    &s^{z}(s^{z})^\ast s^{\frac{z}{2}(\langle q_r\rangle+\langle\alpha_i\rangle)}/ s\rightarrow s;1,\\
		&\sigma_{10}=1,    &s/ s\rightarrow s;1.\\
\\
t_{k+\log_2(z)+3}:\;      &\sigma_1=X+\langle q_r\rangle+\langle\alpha_i\rangle,  &s^{2z}(s^{z})^\ast s^{\langle q_r\rangle+\langle\alpha_i\rangle}/ s^{z}\rightarrow s;4, \\
		&\sigma_4,\sigma_6=1,    &s/s\rightarrow \lambda;0,\\
		&\sigma_7,\sigma_8,\sigma_{9}=z(Y+\langle q_r\rangle+\langle\alpha_i\rangle),    &(s^{z})^\ast/ s\rightarrow \lambda;0,\\
		&\sigma_{10}=z(Y+\langle q_r\rangle+\langle\alpha_i\rangle),    &(s^{z^2})^\ast s^{z(\langle q_r\rangle+\langle\alpha_i\rangle)}/ s^{z^2}\rightarrow s^{z^2};1.\\
\end{xalignat*}
Note that $(zY\mod z^2)=0$ and also that $z(\langle q_r\rangle+\langle\alpha_i\rangle) < z^2$. Thus in neuron $\sigma_{10}$ at time $t_{k+\log_2(z)+3}$ the rule $(s^{z^2})^\ast s^{z(\langle q_r\rangle+\langle\alpha_i\rangle)}/ s^{z^2}\rightarrow s^{z^2};1$ separates the encoding of the right side of the tape $s^{zY}$ and the encoding of the current state and read symbol $s^{z(\langle q_r\rangle+\langle\alpha_i\rangle)}$. To see this note the number of spikes in neurons $\sigma_6$ and $\sigma_{10}$ at time $t_{k+\log_2(z)+4}$.

The rule $s^{z(\langle q_r\rangle+\langle\alpha_i\rangle)}/s^{z(\langle q_r\rangle+\langle\alpha_i\rangle)-\langle q_u\rangle-1}\rightarrow s^{z\langle \alpha_j\rangle};1$, applied in $\sigma_{10}$ at timestep $t_{k+\log_2(z)+4}$, computes the new encoded current state $\langle q_u\rangle$ and the encoded write symbol $z\langle \alpha_j\rangle$. To see this note the number of spikes in neurons $\sigma_6$ and $\sigma_{10}$ at time $t_{k+\log_2(z)+5}$. Note that neuron $\sigma_1$ is preparing to execute the rule $s^{2z}(s^{z})^\ast s^{\langle q_r\rangle+\langle\alpha_i\rangle}/ s^{z}\rightarrow s;1$ at timestep $t_{k+\log_2(z)+6}$, and so at timesteps $t_{k+\log_2(z)+4}$ and $t_{k+\log_2(z)+5}$ neuron $\sigma_1$ remains closed. Thus the spikes sent out from $\sigma_4$ at these times do not enter $\sigma_1$.
\begin{xalignat*}{2}
t_{k+\log_2(z)+4}:\;	&\sigma_1=X+\langle q_r\rangle+\langle\alpha_i\rangle,  &s^{2z}(s^{z})^\ast s^{\langle q_r\rangle+\langle\alpha_i\rangle}/ s^{z}\rightarrow s;3, \\
		&\sigma_4,\sigma_6=zY,    &(s^{z})^\ast/ s\rightarrow s;1,\\
		&\sigma_{10}=z(\langle q_r\rangle+\langle\alpha_i\rangle),   &s^{z(\langle q_r\rangle+\langle\alpha_i\rangle)}/s^{z(\langle q_r\rangle+\langle\alpha_i\rangle)-\langle q_u\rangle-1}\rightarrow s^{z\langle \alpha_j\rangle};1.\\
\\
t_{k+\log_2(z)+5}:\;	&\sigma_1=X+\langle q_r\rangle+\langle\alpha_i\rangle,  &s^{2z}(s^{z})^\ast s^{\langle q_r\rangle+\langle\alpha_i\rangle}/ s^{z}\rightarrow s;2, \\
		&\sigma_2=zY,    & \\
		&\sigma_4,\sigma_6=z\langle \alpha_j\rangle,    &(s^{z})^\ast/ s\rightarrow s;1,\\
		&\sigma_{10}=\langle q_u\rangle+1,   &s^{\langle q_u\rangle+1}/ s^{\langle q_u\rangle}\rightarrow s^{\langle q_u\rangle};4.\\
\\
t_{k+\log_2(z)+6}:\;	&\sigma_1=X+\langle q_r\rangle+\langle\alpha_i\rangle,  &s^{2z}(s^{z})^\ast s^{\langle q_r\rangle+\langle\alpha_i\rangle}/ s^{z}\rightarrow s;1, \\
		&\sigma_2=zY+z\langle \alpha_j\rangle,    & \\
		&\sigma_{10}=\langle q_u\rangle+1,   &s^{\langle q_u\rangle+1}/ s^{\langle q_u\rangle}\rightarrow s^{\langle q_u\rangle};3.
\end{xalignat*}
At time $t_{k+\log_2(z)+7}$ in neuron $\sigma_4$ the rule $s^{z}(s^{z})^\ast s^{(\frac{X}{z} \mod z)} / s^{z}\rightarrow s^{z};1$ is applied sending $\frac{X}{z}-(\frac{X}{z}\mod z)$ spikes to $\sigma_1$ and leaving $(\frac{X}{z}\mod z)$ spikes in $\sigma_4$. At the same time in neuron $\sigma_5$ the rule $s^{z}(s^{z})^\ast s^{(\frac{X}{z} \mod z)} / s^{z}\rightarrow \lambda;0$ is applied leaving only $(\frac{X}{z}\mod z)$ spikes in $\sigma_5$.
\begin{xalignat*}{2}
t_{k+\log_2(z)+7}:\;	&\sigma_1=\langle q_r\rangle+\langle\alpha_i\rangle,  & s^{\langle q_r\rangle+\langle\alpha_i\rangle}/ s\rightarrow \lambda;0, \\
		&\sigma_2=zY+z\langle \alpha_j\rangle,    & \\
		&\sigma_4=\frac{X}{z},    &s^{z}(s^{z})^\ast s^{(\frac{X}{z} \mod z)} / s^{z}\rightarrow s^{z};1, \\
		&\sigma_5=\frac{X}{z},    &s^{z}(s^{z})^\ast s^{(\frac{X}{z} \mod z)} / s^{z}\rightarrow \lambda;0, \\
		&\sigma_{10}=\langle q_u\rangle+1,   &s^{\langle q_u\rangle+1}/ s^{\langle q_u\rangle}\rightarrow s^{\langle q_u\rangle};2.\\
\\
t_{k+\log_2(z)+8}:\;	&\sigma_1=\frac{X}{z}-(\frac{X}{z}\mod z),  & \\
		&\sigma_2=zY+z\langle \alpha_j\rangle,    & \\
		&\sigma_4=\frac{X}{z}\mod z,    &  s^{(\frac{X}{z} \mod z)} / s\rightarrow \lambda;0,\\
		&\sigma_5=\frac{X}{z}\mod z,    & s^{(\frac{X}{z} \mod z)} / s\rightarrow s;1, \\
		&\sigma_{10}=\langle q_u\rangle+1,   &s^{\langle q_u\rangle+1}/ s^{\langle q_u\rangle}\rightarrow s^{\langle q_u\rangle};1.\\
\\
t_{k+\log_2(z)+9}:\;	&\sigma_1=\frac{X}{z}-(\frac{X}{z}\mod z),  & \\
		&\sigma_2=zY+z\langle \alpha_j\rangle,    & \\
		&\sigma_4=\langle q_u\rangle+(\frac{X}{z}\mod z),    & s^{\langle q_u\rangle+(\frac{X}{z}\mod z)}/ s\rightarrow s;1, \\
		&\sigma_6=\langle q_u\rangle+(\frac{X}{z}\mod z),    &s^{\langle q_u\rangle+(\frac{X}{z}\mod z)}/ s\rightarrow s;1,  \\
		&\sigma_7,\sigma_8,\sigma_9=\frac{X}{z} \mod z, &s^{(\frac{X}{z} \mod z)}/s^{(\frac{X}{z} \mod z)}\rightarrow \lambda;0,\\
		&\sigma_{10}=1,    &s/ s\rightarrow s;\log_2(z)+3.
\end{xalignat*}
The simulation of the left moving transition rule is now complete. Note that the number of spikes in $\sigma_1$, $\sigma_2$, $\sigma_4$, and $\sigma_6$ at timestep $t_{k+\log_2(z)+9}$ are the values given by the top case of Equation~\eqref{eq:transition rule simulation} and encode the configuration after the left move transition rule.

The case of when the tape head moves onto a part of the tape that is to the left of $a_{-x+1}$ in Equation~\eqref{eq:configuration} is not covered by the simulation. For example when the tape head is over cell $a_{-x+1}$, then $X=z$ (recall $a_{-x}$ contains $\alpha_1$). If the tape head moves to the left then from the top case of Equation~\eqref{eq:transition rule simulation} the new value for the left sequence is $X=0$. Therefore we increase the length of $X$ to simulate the infinite blank symbols ($\alpha_1$ symbols) to the left as follows. The rule $s^{z+\langle q_r\rangle+\langle\alpha_i\rangle}/s^{z}\rightarrow s^{z};1$ is applied in $\sigma_1$ at time $t_{k+\log_2(z)+6}$. Then at time $t_{k+\log_2(z)+7}$ the rule $(s^{z})^\ast/s\rightarrow s;1$ is applied in $\sigma_4$ and the rule $s^{z}/s^{z-1}\rightarrow \lambda;0$ is applied in $\sigma_5$. Thus at time $t_{k+\log_2(z)+8}$ there are $z$ spikes in $\sigma_1$ which simulates another $\alpha_1$ symbol to the left, and there is 1 spike in $\sigma_5$ to simulate the current read symbol $\alpha_1$.

We have shown how to simulate an arbitrary left moving transition rule~ of~$M$. Right moving transition rules are also simulated in $\log_2(z)+9$ timesteps in a manner similar to that of left moving transition rules. Thus a single transition rule of $M$ is simulated by $\Pi_{M}$ in $\log_2(z)+9$ timesteps. Recall from Section~\ref{sec:Encoding a config of M} ${z={2^{\log_2\lceil 2|Q||A|+2|A|\rceil}}}$ thus the entire computation of $M$ is simulated in $O(|A||Q|T)$ time. From Section~\ref{sec:Encoding a config of M} $M$ is simulated in $O([{2^{\log_2\lceil 2|Q||A|+2|A|\rceil}}]^{T})$ space.
\qed
\end{proof}
While the small universal spiking neural P system in Figure~\ref{fig:universal SNP} simulates Turing machines with a linear time overhead it \emph{requires} an exponential space overhead. This \emph{requirement} may be shown by proving it is simulated by a counter machine using the same space. However, it is not unreasonable to expect efficiency from simple universal systems as many of the simplest computationally universal models have polynomial time and space overheads~\cite{Neary2008,NearyWoods2006C,WoodsNeary2006B}.

It was mentioned in Section~\ref{sec:Spiking neural P-systems} that we generalised the previous definition of spiking neural P systems with exhaustive use of rules to allow the input neuron to receive an arbitrary number of spikes in a single timestep. If the synapses of the system can transmit an arbitrary number of spikes in a single timestep, then it does not seem unreasonable to allow an arbitrary number of spikes to enter the input neuron in a single timestep. If the input is restricted to a constant number of spikes, as is the case with earlier spiking neural P systems, then the system will remain exponentially slow due to the time required to read the unary input into the system.

\begin{table}[h]
\begin{center}
\begin{tabular}{c|@{\quad}l}
neuron       	& rules  \\ \hline
$\sigma_1$		& $(s^{z})^\ast s^{\langle q_r\rangle+\langle\alpha_i\rangle}/ s\rightarrow s;1$   if D=R\\ 
			& $s^{2z}(s^{z})^\ast s^{\langle q_r\rangle+\langle\alpha_i\rangle}/ s^{z}\rightarrow s;\log_2(z)+6$ if D=L \\ 
			& $ s^{z+\langle q_r\rangle+\langle\alpha_i\rangle}/ s^{z}\rightarrow s^{z};\log_2(z)+6$ if D=L\\ 
			& $s^{\langle q_r\rangle+\langle\alpha_i\rangle}/ s\rightarrow \lambda;0$ if D=L\\ 
\hline 
$\sigma_2$		& $(s^{z})^\ast s^{\langle q_r\rangle+\langle\alpha_i\rangle}/ s\rightarrow s;1$   if D=L or $\langle q_r\rangle=\langle q_{|Q|}\rangle$\\ 				& $s^{2z}(s^{z})^\ast s^{\langle q_r\rangle+\langle\alpha_i\rangle}/ s^{z}\rightarrow s;\log_2(z)+6$ if D=R \\ 
			& $ s^{z+\langle q_r\rangle+\langle\alpha_i\rangle}/ s^{z}\rightarrow s^{z};\log_2(z)+6$ if D=R\\ 
			& $s^{\langle q_r\rangle+\langle\alpha_i\rangle}/ s\rightarrow \lambda;0$ if D=R\\ 
\hline 
$\sigma_3$   		& $(s^{z})^\ast s^{\langle q_r\rangle+\langle\alpha_i\rangle}/ s^{z}\rightarrow s^{z};1,$ if $\langle q_r\rangle=\langle q_{|Q|}\rangle$\\
			& $(s^{z})^\ast s^{\langle q_r\rangle+\langle\alpha_i\rangle}/ s\rightarrow \lambda;0,$ if $\langle q_r\rangle\neq\langle q_{|Q|}\rangle$\\
			& $(s^{z})^\ast s^{(\frac{X}{z} \mod z)} / s\rightarrow \lambda;0$\\
			& $s^{z} / s\rightarrow \lambda;0$\\ 
\hline
$\sigma_4$		& $s^{2}(s^{z})^\ast /s^{z}\rightarrow s^{z};2$ \\
			& $(s^{z})^\ast/ s\rightarrow s;1$ \\ 
			& $s^{2} /s^2\rightarrow \lambda;0$ \\ 
			& $s^{\langle q_r\rangle+\langle\alpha_i\rangle}/ s\rightarrow s;1$ \\
			& $s^{z}(s^{z})^\ast s^{\langle q_r\rangle+\langle\alpha_i\rangle}/ s\rightarrow \lambda;0$ \\
			& $s/s\rightarrow \lambda;0$ \\ 
			& $s^{z}(s^{z})^\ast s^{(\frac{X}{z} \mod z)} / s^{z}\rightarrow s^{z};1$ \\ 
			& $s^{(\frac{X}{z} \mod z)} / s\rightarrow \lambda;0$ \\ 
\hline 
$\sigma_5$		& $s^{2}(s^{z})^\ast /s\rightarrow s;1$ \\ 
			& $s^{2z}(s^{z})^\ast /s\rightarrow s;1$ \\
			& $(s^{z})^\ast s^{\langle q_r\rangle+\langle\alpha_i\rangle}/ s\rightarrow s;1$ \\
			& $s^{z}(s^{z})^\ast s^{(\frac{X}{z} \mod z)} / s^{z}\rightarrow \lambda;0$ \\
			& $s^{z} /s^{z-1}\rightarrow \lambda;0$ \\
			& $s^{(\frac{X}{z} \mod z)} / s\rightarrow s;1$ \\
\hline
$\sigma_6$		& $s^{2}(s^{z})^\ast /s\rightarrow \lambda;0$ \\ 
			& $(s^{z})^\ast/ s\rightarrow s;1$ \\ 
			& $s^{\langle q_r\rangle+\langle\alpha_i\rangle}/ s\rightarrow s;1$ \\
			& $s^{z}(s^{z})^\ast s^{\langle q_r\rangle+\langle\alpha_i\rangle}/ s\rightarrow \lambda;0$ \\
			& $s/s\rightarrow \lambda;0$ \\ 
			& $s^{z}(s^{z})^\ast s^{(\frac{Y}{z} \mod z)} / s^{z}\rightarrow s^{z};1$ \\ 
			& $s^{(\frac{Y}{z} \mod z)} / s\rightarrow \lambda;0$ \\ 
\hline 
\end{tabular}
\end{center}
\caption{This table gives the rules in each of the neurons $\sigma_1$ to $\sigma_6$ of $\Pi_{M}$. In the rules above $q_r$ is the current state, $\alpha_i$ is the read symbol, $\alpha_j$ is the write symbol, $D$ is the move direction, and $q_u$ is the next state of some transition rule
$q_r,\alpha_i,\alpha_j,D,q_u$ of $M$. Note that $(\frac{X}{z} \mod z)),(\frac{Y}{z} \mod z))\in \langle A\rangle$ the set of encodings for the symbols of $M$ (see Section~\ref{sec:Encoding a config of M}).}\label{tab:neurons of universal SNP I}
\end{table}

\begin{table}[h]
\begin{center}
\begin{tabular}{c|@{\quad}l}
neuron       	& rules  \\ \hline
$\sigma_7,\sigma_8,\sigma_9$	& $s^{2}(s^{z})^\ast /s\rightarrow \lambda;0$ \\
				& $(s^{z})^\ast /s\rightarrow \lambda;0$ \\ 
				& $s^{\langle q_r\rangle+\langle \alpha_i\rangle} /s\rightarrow \lambda;0$ \\
				& $s^{z}(s^{z})^\ast s^{\frac{z}{m}(\langle q_r\rangle+\langle\alpha_i\rangle)}/ s\rightarrow s;1$\;\; for all $m=2^k$, $2\leqslant m\leqslant z$  and $k\in\Nset$\\
				& $s^{(\frac{X}{z} \mod z)}/s^{(\frac{X}{z} \mod z)}\rightarrow \lambda;0$ \\
\hline
$\sigma_{10}$		   	& $s^{31}/ s^{16}\rightarrow \lambda;0$ \\
				& $s^{15}/ s^{8}\rightarrow \lambda;0$ \\
				& $s^{7}/ s^{4}\rightarrow \lambda;0$ \\
				& $s^{3}/ s^{2}\rightarrow \lambda;0$ \\
				& $s/ s\rightarrow s;\log_2(z)+3$ \\
				& $(s^{z^2})^\ast s^{z(\langle q_r\rangle+\langle\alpha_i\rangle)}/ s^{z^2}\rightarrow s^{z^2};1$ \\
				& $s^{z(\langle q_r\rangle+\langle\alpha_i\rangle)}/s^{z(\langle q_r\rangle+\langle\alpha_i\rangle)-\langle q_u\rangle-1}\rightarrow s^{z\langle \alpha_j\rangle};1$ \\
				& $s^{\langle q_u\rangle+1}/ s^{\langle q_u\rangle}\rightarrow s^{\langle q_u\rangle};4$ \\ 
\hline 
\end{tabular}
\end{center}
\caption{This table gives the rules in each of the neurons $\sigma_7$ to $\sigma_{10}$ of $\Pi_{M}$. See Table~\ref{tab:neurons of universal SNP I} for some further explanation.}\label{tab:neurons of Extended SNP II}
\end{table}

\begin{table}[h]
\begin{center}
\begin{tabular}{c|@{\quad}l}
neuron       	& rules  \\ \hline
$\sigma_1$			& $s^\ast /s\rightarrow s;1$ \\
\hline
$\sigma_2,\sigma_3,\sigma_4$	& $s^\ast/s\rightarrow s;1$ \\
\hline
$\sigma_{5}$		   	& $(s^{z})^\ast s^{\langle \alpha\rangle}/s\rightarrow s;\log_2(z)$ \\
				& $(s^{z})^\ast s^2/s\rightarrow s;1$ \\
\hline
$\sigma_{6}$		   	& $(s^{z})^\ast s^{\langle a\rangle}/ s\rightarrow \lambda;0$ \\
				& $(s^{z})^\ast s^2/ s^{z}\rightarrow s^{z};1$ \\
\hline 
\end{tabular}
\end{center}
\caption{This table gives the rules in each of the neurons of $\Pi_{input}$.}\label{tab:rules for input SNP}
\end{table}


\begin{thebibliography}{10}

\bibitem{Chen2006A}
H.~Chen, M.~Ionescu, and T.~Ishdorj.
\newblock On the efficiency of spiking neural {P} systems.
\newblock In {M.A. Guti\'{e}́rrez-Naranjo et al.}, editor, {\em Proceedings of
  Fourth Brainstorming Week on Membrane Computing}, pages 195--206, Sevilla,
  Feb. 2006.

\bibitem{Fischer1968}
P.~C. Fischer, A.~Meyer, and A.~Rosenberg.
\newblock Counter machines and counter languages.
\newblock {\em Mathematical Systems Theory}, 2(3):265--283, 1968.

\bibitem{Ionescu2006}
M.~Ionescu, G.~P{\u{a}}un, and T.~Yokomori.
\newblock Spiking neural {P} systems.
\newblock {\em Fundamenta Informaticae}, 71(2-3):279--308, 2006.

\bibitem{Ionescu2007B}
M.~Ionescu, G.~P{\u{a}}un, and T.~Yokomori.
\newblock Spiking neural {P} systems with exhaustive use of rules.
\newblock {\em International Journal of Unconventional Computing},
  3(2):135--153, 2007.

\bibitem{Ionescu2007A}
M.~Ionescu and D.~Sburlan.
\newblock Some applications of spiking neural {P} systems.
\newblock In {George Eleftherakis et al.}, editor, {\em Proceedings of the
  Eighth Workshop on Membrane Computing}, pages 383--394, Thessaloniki, June
  2007.

\bibitem{Korec1996}
I.~Korec.
\newblock Small universal register machines.
\newblock {\em Theoretical Computer Science}, 168(2):267--301, Nov. 1996.

\bibitem{Leporati2007A}
A.~Leporati, C.~Zandron, C.~Ferretti, and G.~Mauri.
\newblock On the computational power of spiking neural {P} systems.
\newblock In {M.A. Guti\'{e}́rrez-Naranjo et al.}, editor, {\em Proceedings of
  the Fifth Brainstorming Week on Membrane Computing}, pages 227--245, Sevilla,
  Jan. 2007.

\bibitem{Leporati2007B}
A.~Leporati, C.~Zandron, C.~Ferretti, and G.~Mauri.
\newblock Solving numerical {NP}-complete problems with spiking neural {P}
  systems.
\newblock In {George Eleftherakis et al.}, editor, {\em Proceedings of the
  Eighth Workshop on Membrane Computing}, pages 405--423, Thessaloniki, June
  2007.

\bibitem{Neary2009}
T.~Neary.
\newblock A boundary between universality and non-universality in spiking neural P systems.
\newblock arXiv:0912.0741v1 [cs.CC]. December 2009.

\bibitem{Neary2008A}
T.~Neary.
\newblock On the computational complexity of spiking neural {P} systems.
\newblock In {\em Unconventional Computation, 7th International Conference, UC
  2008}, volume 5204 of {\em LNCS}, pages 189--205, Vienna, Aug. 2008.
  Springer.

\bibitem{Neary2008B}
T.~Neary.
\newblock A small universal spiking neural {P} system.
\newblock In {\em International Workshop on Computing with Biomolecules}, pages
  65--74, Vienna, Aug. 2008. Austrian Computer Society.

\bibitem{Neary2008C}
T.~Neary.
\newblock Presentation at The International Workshop on Computing with
  Biomolecules (CBM 2008). Available at
  http://www.emcc.at/UC2008/Presentations/CBM5.pdf\;.

\bibitem{Neary2008}
T.~Neary.
\newblock {\em Small universal {T}uring machines}.
\newblock PhD thesis, National University of Ireland, Maynooth, Oct. 2008.

\bibitem{NearyWoods2006C}
T.~Neary and D.~Woods.
\newblock {$\P$}-completeness of cellular automaton {R}ule 110.
\newblock In {Michele Bugliesi et al.}, editor, {\em International Colloquium
  on Automata Languages and Programing 2006, (ICALP) Part {I}}, volume 4051 of
  {\em LNCS}, pages 132--143, Venice, July 2006. Springer.

\bibitem{Paun2007}
A.~P{\u{a}}un and G.~P{\u{a}}un.
\newblock Small universal spiking neural {P} systems.
\newblock {\em BioSystems}, 90(1):48--60, 2007.

\bibitem{Paun2002}
G.~P{\u{a}}un.
\newblock {\em Membrane Computing: An Introduction}.
\newblock Springer, 2002.

\bibitem{WoodsNeary2006B}
D.~Woods and T.~Neary.
\newblock On the time complexity of 2-tag systems and small universal {T}uring
  machines.
\newblock In {\em 47$^{\textrm{th}}$ Annual IEEE Symposium on Foundations of
  Computer Science (FOCS)}, pages 439--448, Berkeley, California, Oct. 2006.
  IEEE.

\bibitem{Zhang2008A}
X.~Zhang, Y.~Jiang, and L.~Pan.
\newblock Small universal spiking neural {P} systems with exhaustive use of
  rules.
\newblock In {\em 3rd International Conference on Bio-Inspired Computing:
  Theories and Applications(BICTA 2008)}, pages 117--128, Adelaide, Australia,
  Oct. 2008. IEEE.

\bibitem{Zhang2008B}
X.~Zhang, X.~Zeng, and L.~Pan.
\newblock Smaller universal spiking neural {P} systems.
\newblock {\em Fundamenta Informaticae}, 87(1):117--136, Nov. 2008.

\end{thebibliography}
\end{document}